\documentclass[sigconf]{acmart}

\AtBeginDocument{%
  \providecommand\BibTeX{{%
    \normalfont B\kern-0.5em{\scshape i\kern-0.25em b}\kern-0.8em\TeX}}}
    
\usepackage{subcaption}
\usepackage{graphicx}
\usepackage{caption}
\usepackage{xcolor}
\usepackage{amsmath}
\usepackage{titlesec}
\usepackage{algorithm}
\usepackage{algpseudocode}
\usepackage{threeparttable}
\usepackage{multirow}
\usepackage{tabularx}

\titlespacing*{\section}{0pt}{5pt plus 1pt}{5pt minus 1pt}
\titlespacing*{\subsection}{0pt}{3pt plus 1pt}{3pt minus 1pt}
\titlespacing*{\subsubsection}{0pt}{2pt plus 1pt}{2pt minus 1pt}

\begin{document}

\title{Learning to Rank Rationales for Explainable Recommendation}

\author{Zhichao Xu}
\email{zhichao.xu@utah.edu}
\affiliation{%
  \institution{University of Utah}
  \country{United States}
}

\author{Yi Han}
\email{yi.han@rutgers.edu}
\affiliation{%
  \institution{Rutgers University}
  \country{United States}
}

\author{Tao Yang}
\email{taoyang@cs.utah.edu}
\affiliation{%
  \institution{University of Utah}
  \country{United States}
}

\author{Anh Tran}
\email{abtran@cs.utah.edu}
\affiliation{%
  \institution{University of Utah}
  \country{United States}
}

\author{Qingyao Ai}
\email{aiqy@cs.utah.edu}
\affiliation{%
  \institution{University of Utah}
  \country{United States}
}

\begin{abstract}
State-of-the-art recommender system (RS) mostly rely on complex deep neural network (DNN) model structure, which makes it difficult to provide explanations along with RS decisions. 
Previous researchers have prove that providing explanations with recommended items can help users make informed decisions and improve users' trust towards the uninterpretable blackbox system.
In model-agnostic explainable recommendation, system designers deploy a separate explanation model to take the DNN model's decision as input, and generate explanations to meet the goal of persuasiveness.
In this work, we explore the task of ranking textual rationales (supporting evidences) for model-agnostic explainable recommendation. 
Most of existing rationales ranking algorithms only utilize the rationale IDs and interaction matrices to build latent factor representations; and the semantic information within the textual rationales are not learned effectively. 
We argue that such design is suboptimal as the important semantic information within the textual rationales may be used to better profile user preferences and item features.
Seeing this gap, we propose a model named \textbf{S}emantic-\textbf{E}nhanced \textbf{B}ayesian \textbf{P}ersonalized \textbf{E}xplanation \textbf{R}anking (\textbf{SE-BPER}) to effectively combine the interaction information and semantic information.
SE-BPER first initializes the latent factor representations with contextualized embeddings generated by transformer model, then optimizes them with the interaction data.
Extensive experiments show that such methodology improves the rationales ranking performance while simplifying the model training process (fewer hyperparameters and faster convergence).
We further design a pretraining scheme to finetune a transformer model for better latent factors initialization. 
We conclude that the optimal way to combine semantic and interaction information remains an open question in the task of rationales ranking.

\end{abstract}


\maketitle

\section{Introduction}
With the emerging development of Internet in the past few decades, researchers have devoted extensive effects to the study of recommender system (RS) \cite{koren2008factorization,koren2009matrix,mcauley2013hidden,xu2020commerce,yang2022effective}.
Existing real-world RS applications mostly adopt complex deep neural network (DNN) structures. 
While complex network structures help with model performance, this also leads to the loss of model interpretability.
A recent line of works focus on the explainability of recommender system \cite{zhang2018learning,ai2018learning,pena2020combining,chen2018neural,zeng2021zero} to improve the transparency \cite{jacovi2020towards}, persuasiveness \cite{li2021learning,ai2018learning,ai2021model}, trustworthiness \cite{jacovi2021formalizing}, as well as user satisfaction \cite{ai2021model} of the existing applications \cite{zhang2020explainable}. 

Existing works on explainable recommendation can be broadly classified into two categories: \textit{model-intrinsic} (or pre-hoc) explanations and \textit{model-agnostic} (or post-hoc) explanations. 
In model-instrinsic explanation paradigm, researchers aim to design interpretable model structures to generate explanations alongside model decisions \cite{chen2018neural,xian2019reinforcement,xian2020cafe}; while in model-agnostic explanation paradigm, the explanations are generated from a separate explainer, to meet certain goals, such as \textit{faithfulness}, \textit{comprehensiveness}, \textit{persuasiveness} or \textit{plausibility}. 
In recommendation, \textit{persuasiveness} is defined as the ability to persuade the user into interacting with the item and making informed decisions that are supported by evidences \cite{schafer2001commerce,zhang2020explainable}. 
Within the scope of this work, we mainly focus on improving the persuasiveness of recommendation explanations.
Specifically, Li et al. \cite{li2021extra,li2021learning} first propose a task of ranking rationales for improving the persuasiveness of explainable recommendation.
The task is defined as: given user $u$ and item $i$ recommended by the recommendation model, 
to design an explainer to select the most accurate textual rationales from a large collection to explain and persuade the users of the item's relevance to their needs.
In this case, the explainer is a separate model from the decision model, thus we consider the task of rationales ranking as a subtask under the model-agnostic explanation paradigm.

The representative algorithm for rationales ranking Bayesian Personalized Explanation Ranking (BPER) \cite{li2021learning} only utilizes the IDs of the users, items and rationales and trains their latent factor representations by performing matrix factorization over the interaction matrices. 
While efficient, this algorithm fails to catch the rich semantic information in the textual rationales, which can be critical for the task of rationales ranking.
In the same work, the authors \cite{li2021learning} also propose to incorporate the semantic information into latent factor representations. The combination is performed by multiplying the semantic latent factors with the vanilla latent factors.
The semantic latent factors are derived by linearly projecting the contextualized representations onto a smaller latent space.
We argue this method may lead to loss of performance in the projection phase. 
Also, the initialization of semantic latent factors are derived from a static transformer model, and remain static during training, which may not catch the task-specific semantic signals.


Seeing this gap, we propose \textbf{SE-BPER} (short for \textbf{S}emantic-\textbf{E}nhanced \textbf{B}ayesian \textbf{P}ersonalized \textbf{E}xplanation \textbf{R}anking) algorithm for the task of rationales ranking for explainable recommendation.
SE-BPER replaces the random latent factors initialization of BPER with semantic representations and achieves significantly better performances validated by extensive experiments.
We further propose a task-specific pretrain scheme to improve the semantic initialization by finetuning a transformer model on rationales ranking dataset, which we term as \textbf{SE-BPER+}.
Our contributions can be summarized as follows:
\begin{itemize}
    \vspace{-2pt}
    \item We study the task of rationales ranking under the model-agnostic explanation paradigm and propose \textbf{SE-BPER} to effectively capture the rich semantic information in textual rationales.
    \item We further propose a task-specific pretrain scheme to enhance the learning of semantic information by finetuning a transformer model on the rationales ranking dataset, which we term as \textbf{SE-BPER+}
    \item Extensive experiments show that our SE-BPER achieves significant improvement over state-of-the-art algorithm BPER. Our finetuning scheme SE-BPER+ can further boost the performance.
\end{itemize}
The rest of the paper is organized as follows: we review the related work in Sec \ref{sec:related}.
Then in Sec \ref{sec:prior} we introduce the task of rationales ranking and analyze the weakness of BPER, the current SOTA algorithm for this task.
In Sec \ref{sec:model} we introduce the proposed SE-BPER and SE-BPER+ algorithm.
The experimental settings and details are covered in Sec \ref{sec:experiment}. 
We report the experiment results and conduct analysis in Sec \ref{sec:result}.
Finally we conclude this work in Sec \ref{sec:conclusion}.
Our implementation is public\footnote{https://github.com/zhichaoxu-shufe/Semantic-Enhanced-BPER}.

\section{Related Work}
\label{sec:related}
There are three lines of work that are related to our work: Explainable AI and Recommendation, Model-agnostic Explainable Recommendation and Initialization in Latent Factor Models.

\textbf{\textit{Explainable AI and Recommendation}}. 
Explainable AI aims to provide explanations along with AI system's decisions to improve the users' overall satisfaction.
Existing explainable AI research can be broadly divided into two directions: \textit{model-intrinsic} and \textit{model-agnostic}. 
In model-intrinsic explanation paradigm, researchers aim to build interpretable model structures to generate explanations alongside model decisions \cite{ai2018learning,chen2018neural,bahdanau2014neural}; a representative line of works is the attention mechanism \cite{bahdanau2014neural,chen2018neural,chen2019dynamic}. 
In model-agnostic explanation paradigm, the aim is to design a separate explainer to explain the decision model's behaviors to meet certain objectives, such as \textit{faithfulness}, \textit{scrutability}, \textit{comprehensiveness}, \textit{persuasiveness} and \textit{plausibility}.
In the scope of explainable recommendation, \textit{persuasiveness} is defined as the ability to persuade the user into interacting with the item and making informed decisions that are supported by evidences \cite{schafer2001commerce,zhang2020explainable}. 
Within the scope of this work, we mainly focus on improving the persuasiveness of the recommendation explanations. 

The evaluation of Explainable Recommendation remains an open question. 
Some of existing works adopt user study \cite{ai2021model,ghazimatin2020prince} to prove their methods are human-interpretable and can improve persuasiveness of the recommender system\footnote{also referred to as \textit{human-grounded evaluation} by \cite{madsen2021post}}. 
However, there hasn't been a universal way to quantitatively evaluate the quality of explanations, and the metrics adopted are mostly task and model specific.
Within the scope of this work, the explanation task is formulated as a ranking problem, where the groundtruth rationales are pre-extracted from the user-generated product reviews \cite{mcauley2013hidden,pena2020combining,zeng2021zero,xu2021understanding}\footnote{also referred to as \textit{Extractive Rationales} by \cite{lei2016rationalizing} and \textit{Highlights} by \cite{wiegreffe2021teach}}.
Under this formulation, we could use the classical ranking metrics such as Recall and nDCG to quantitatively evaluate the effectiveness of the proposed model.

\textbf{\textit{Model-agnostic Explainable Recommendation}}. 
Depending on the specific problem settings, the model-agnostic explainable recommendation algorithms can be further divided into those designed for black-box models \cite{nobrega2019towards,peake2018explanation} (neither the model structure nor the parameters/gradients are known by the explainer) and those designed for grey-box models \cite{cheng2019incorporating,tran2021counterfactual} (the gradients or the embeddings of the recommendation model are available to the explanation model)\footnote{interested readers can refer to \cite{madsen2021post} for a more detailed taxonomy.}.
There have been some works on improving the effectiveness of model-agnostic explainable recommendation paradigm. 
Peake and Wang \cite{peake2018explanation} use a different explainer to explain latent factor recommendation models.
Ai et al. \cite{ai2018learning} propose a surrogate explanation model using knowledge graph paths as explanations. 
In a more recent work, Ai and Narayanan \cite{ai2021model} compare model-agnostic explanation with model-intrinsic explanation from the perspective of persuasiveness and plausibility.
Within the scope of this work, we only focus on designing an explainer, which takes the user-item pair as input and will select some rationales as explanations.
Under this setup, our explainer falls into the scope of black-box model, as it has no knowledge about the recommendation model other than the recommended item\footnote{The explainer will still have access to the user and item rationales histories, but these information are not necessarily available for the recommender model.}.

\textbf{\textit{Initialization in Latent Factor Models}}. 
Traditionally, matrix factorization (MF) models are initialized with random values \cite{hidasi2013initializing,albright2006algorithms,sun2020we}. 
Previous researhers also show that with carefully-designed initialization strategies, MF models can achieve faster convergence and reach better performance \cite{boutsidis2008svd,hidasi2012enhancing,pena2020combining}.
Pena et al. \cite{pena2020combining} propose to initialize the user/item latent factors with pretrained embeddings from topic models and deliver better top-\textit{N} recommendation performance. 
Our work is inspired by \cite{pena2020combining} but we have two key differences: 
(1) we explore the task of ranking rationales, where users, items and rationales form a triparitite graph rather than user-item bipartite graph; 
(2) the embeddings in the topic space in \cite{pena2020combining} are trained with only nouns in the vocabulary, while we leverage a powerful pretrained contextualized language model to obtain better representations.

\section{Prior Knowledge}
\label{sec:prior}

\begin{table}[t]
\caption{A summary of notations}
\resizebox{0.85\columnwidth}{!}{
\begin{tabular}{l|l}
\toprule
$u$, $\mathcal{U}$ & user, user set\\
$i$, $\mathcal{I}$ & item, item set\\
$e$, $\mathcal{E}$ & rationale, rationales set\\
$\mathcal{E}_u$ & the rationales $u$ mentioned in trainset\\
$\mathcal{E}_i$ & the rationales $i$ was mentioned on in trainset\\
$\mathcal{D}$ & the decision model \\
$U$ & the user-rationale interaction matrix\\
$I$ & the item-rationale interaction matrix\\
$\textbf{p}_u$ & the user latent factor for $U$\\
$\textbf{o}_e^U$ & the rationale latent factor for $U$\\
$\textbf{q}_i$ & the item latent factor for $I$\\
$\textbf{o}_e^I$ & the rationale latent factor for $I$\\
$b_e^U$ & the rationale bias for $U$\\
$b_e^I$ & the rationale bias for $I$\\
$r_{ue}$ & user $u$'s preference over rationale $e$ \\
$r_{ie}$ & item $i$'s appropriateness for rationale $e$ \\
$r_{uie}$ & the score of $(u,i)$ over rationale $e$ \\

\bottomrule
\end{tabular}
}
\label{tab:notations}
\end{table}

We first introduce the rationales ranking task and why it falls into the scope of the model-agnostic explainable recommendation. 
Then we talk about the problem formulation and its notations. 
After that we conduct a comprehensive review of Bayesian Personalized Explanation Ranking (BPER) \cite{li2021learning}, the state-of-the-art model for rationales ranking, and its improved version BPER+.
We go further to discuss why BPER and BPER+ are suboptimal in incorporating semantic information from textual rationales.

\subsection{Rationales Ranking Task}
In Explainable AI literature, the rationales are defined as the supporting evidences for the model decisions and can be flexible in terms of forms such as annotated text highlights \cite{lamm2020qed}, visual reasoning graphs \cite{marasovic2020natural}, counterfactuals \cite{ross2020explaining} and important pixels \cite{ribeiro2016should}. 
In the scope of explainable recommendation, Li et al. \cite{li2021extra} extract the short textual snippets from user-generated textual reviews and use those as the ground truth rationales explaining why the item might be relevant to the user.

Under this setup, the rationales ranking task also falls into the scope of \textit{model-agnostic} explainable recommendation.
The recommendation decision of recommending item $i$ to user $u$ is generated by a recommendation model,
and a separate explainer will select some rationales from a large collection of candidate rationales to explain why this item might be relevant to the user to achieve the \textit{persuasiveness}.
In this case, the model-agnostic explanation paradigm doesn't rely on the specific choice of the decision model\footnote{the decision model is treated as a blackbox} and only focuses on how to persuade a user of the relevance of a specific item.

\subsection{Problem Formulation and Notations}
\label{sec:notations}
The rationales are defined as the supporting evidences in the form of short textual snippets explaining the reason why item $i$ might be relevant to the user $u$, such as \textit{Excellent movie} or \textit{Wonderful movie for all ages}. 
The candidate rationales are extracted from the the user generated product reviews, where the detailed extraction algorithm is introduced in \cite{li2021extra}.

Denote the users set, items set, rationales set as $\mathcal{U}$, $\mathcal{I}$, $\mathcal{E}$, respectively; 
and user-rationale interaction matrix $U$ records the historical interactions between $\mathcal{U}$ and $\mathcal{E}$, 
similarly, item-rationale interaction matrix $I$ records the interactions between $\mathcal{I}$ and $\mathcal{E}$.
Formally, the task is defined as: given user $u \in \mathcal{U}$  and item $i \in \mathcal{I}$ recommended by the Oracle decision model $\mathcal{D}$, to rank the candidate rationales $e \in \mathcal{E}$ according to the estimated relevance $\hat{r}_{uie}$ between $(u,i)$ pair and the rationale $e$. 
In this work, we assume that the recommended item $i$ is already given by a so-called Oracle model $\mathcal{D}$ and only focus on the task of rationales ranking.
We show a summary of notations in Table \ref{tab:notations}.

\subsection{BPER and BPER+}
\label{sec:bper}
Given that the users, items and rationales form a tripartite graph, 
Li et al. \cite{li2021extra,li2021learning} propose to train the user, item and rationale embeddings using the classical tensor decomposition algorithm.
The idea is to extend matrix factorization on interaction matrix to tensor decomposition on the interaction cube. 
Li et al. \cite{li2021learning} show this approach can't provide satisfying performance, as the interaction cube of shape $\mathcal{U}\times \mathcal{I} \times \mathcal{E}$ is highly sparse and the latent factor embeddings can't be trained sufficiently. 
\subsubsection{BPER model.}
\label{subsec:bper}
Li et al. \cite{li2021learning} propose BPER to mitigate this sparsity problem. The idea of BPER is to perform two sets of matrix factorization on user-rationale interaction matrix $U$ and item-rationale interaction matrix $I$ simultaneously. 
More specifically, the estimated preferences on $U$, $I$ are modeled by
\begin{equation}
    \hat{r}_{ue} = \langle \textbf{p}_u^\text{T}, \textbf{o}_e^U \rangle + b_e^U
    \label{eq:rue}
\end{equation}
\begin{equation}
    \hat{r}_{ie} = \langle \textbf{q}_i^\text{T}, \textbf{o}_e^I \rangle + b_i^U
    \label{eq:rie}
\end{equation}
respectively, where $\hat{r}_{ue}$ denotes the estimated user $u$'s preference over rationale $e$, $\hat{r}_{ie}$ denotes the estimated item $i$'s appropriateness for rationale $e$, and $\langle \cdot, \cdot \rangle$ denotes dot product; The authors propose to learn the pairwise preference from $U$ and $I$, and we have
\begin{equation}
    \hat{r}_{u,ee'} = \hat{r}_{ue}-\hat{r}_{ue'}
\end{equation}
\begin{equation}
    \hat{r}_{i,ee''} = \hat{r}_{ie}-\hat{r}_{ie''}
\end{equation}
which reflect user $u$'s preference in rationale $e$ over $e'$, and item $i$'s appropriateness for rationale $e$ over $e''$, respectively.
BPER is optimized with a BPR loss \cite{rendle2012bpr} by minimizing:
\begin{equation}
    \mathcal{L} = \sum_{u \in \mathcal{U}} \sum_{i \in \mathcal{I}} \sum_{e \in \mathcal{E}} ( -\sum_{e'} \ln \sigma(\hat{r}_{u,ee'}) - \sum_{e''} \ln \sigma(\hat{r}_{i,ee''}) )
    \label{eq:loss}
\end{equation}
where $e'$ and $e''$ are negative samples; 
$e' \in \Omega(\mathcal{E}\backslash \mathcal{E}_u)$ and $e''  \in \Omega(\mathcal{E}\backslash \mathcal{E}_i)$, and $\Omega(\cdot)$ denotes the negative sampling procedure for noise contrastive loss similar to \cite{mikolov2013distributed,ai2018learning,zhang2021understanding}.

\subsubsection{BPER inference.}
After training the user, item and rationale embeddings by Eq. \ref{eq:loss}, the estimated preference of $(u,i)$ pair on rationale $e$ is calculated by 
\begin{equation}
    \hat{r}_{uie} = \mu \cdot \hat{r}_{ue} + (1-\mu) \cdot \hat{r}_{ie}
    \label{eq:score}
\end{equation}
where $\hat{r}_{ue}$ and $\hat{r}_{ie}$ are derived from Eq. \ref{eq:rue} and Eq. \ref{eq:rie}, respectively.
For each $(u,i)$ pair in the testset, the ranklist is constructed by sorting the rationales according to their estimated preference scores.

\subsubsection{BPER+ model.}
\label{subsec:bper+}
BPER+ takes one step further to combine the semantic information of the textual rationales. Similar to \cite{nogueira2019bert,macavaney2019cedr}, the authors first input the textual rationale into a pretrained transformer-based language model (BERT \cite{devlin2018bert} in this case), and use the output \text{[CLS]} token's embedding as the semantic representation of the textual rationale, which has the latent dimension of 768.
The authors then perform a linear projection to project the semantic embedding into the same dimension as the vanilla rationale embedding,
\begin{equation}
\textbf{o}_e^{\textit{BERT}} = W_{d \times 768} * \textbf{e}_{\text{[CLS]}}
\label{eq:projection}
\end{equation}
where $d$ is the size of vanilla rationale embedding, $W_{d \times 768}$ is the projection matrix, $\textbf{e}_{\text{[CLS]}}$ is the 768-dimension vector of the [CLS] token, and $*$ denotes matrix multiplication. The enhanced rationale vectors $\textbf{o}_e^{U+}$ and $\textbf{o}_e^{I+}$ are derived by multiplying $\textbf{o}_e^{\textit{BERT}}$
\begin{equation}
    \textbf{o}_e^{U+}=\textbf{o}_e^{U} \odot \textbf{o}_e^{\textit{BERT}}
    \label{eq: u_mul}
\end{equation}
\begin{equation}
    \textbf{o}_e^{I+}=\textbf{o}_e^{I} \odot \textbf{o}_e^{\textit{BERT}}
    \label{eq: i_mul}
\end{equation}
where $\odot$ denotes element-wise multiplication. And the rest parts of BERT+ are the same as BPER.


\subsubsection{Analysis of BPER.} 
From Sec \ref{subsec:bper} we can see BPER doesn't make use of the semantic information in the textual rationales. 
The user, item and rationale embeddings are optimized using the implicit feedback interaction matrices.
We argue this design is suboptimal as textual rationales may contain important semantic information to profile users and items.

BPER+ propose to incorporate the semantic information by first projecting the contextualized representations into the same dimension as the vanilla rationale embeddings, then combining the two embeddings via multiplication. 
We argue there are two problems with this design: (1) in the projection phase, important semantic information may be lost; (2) the initialization of semantic latent factors are generated by a static transformer model, and remains static during training, which may not catch the task-specific semantic signals.

\section{The Proposed Algorithms}
Comprehensive analysis demonstrates that BPER and BPER+ still can't incorporate the semantic information effectively, thus we propose \textbf{S}emantic-\textbf{E}nhanced \textbf{B}ayesian \textbf{P}ersonalized \textbf{E}xplanation \textbf{R}anking, short for \textbf{SE-BPER}.
\label{sec:model}
\subsection{SE-BPER}
Our SE-BPER can be seen as an improved version of BPER+. Instead of performing the burdensome projection and multiplication operations, we directly initialize $\textbf{p}_u$, $\textbf{p}_i$, $\textbf{o}_e^U$, $\textbf{o}_e^I$ with the semantic embeddings.
The intuition is that the semantic information learned by the transformer model can serve as a good initialization for matrix factorization; and helps with model convergence. 
More specifically, we have 
\begin{equation}
    \textbf{o}_e^{U} = \textbf{o}_e^{I} = \textbf{e}_{\text{[CLS]}}
\end{equation}
The user latent factor is derived by averaging the embeddings of all rationales mentioned by the user $u$ in the trainset:
\begin{equation}
    \textbf{p}_u = \sum_{e \in \mathcal{E}_u} \textbf{e}_{\text{[CLS]}} / |\mathcal{E}_u|
    \label{eq:p_u}
\end{equation}
where $|\cdot|$ denotes the size of the set. And similarily,
\begin{equation}
    \textbf{q}_i = \sum_{e \in \mathcal{E}_i} \textbf{e}_{\text{[CLS]}} / |\mathcal{E}_i|
    \label{eq:q_i}
\end{equation}
Note that here we take a simple averaging operation here, and the design can be further improved by more complex aggregation function, such as weighted averaging or attention.

Our SE-BPER is conceptually better than BPER+ as it avoids the loss of information in the projection phase in Equation \ref{eq:projection}. However, the semantic information still remains static during training, thus we propose SE-BPER+.

\subsection{SE-BPER+}
\label{sec:se-bper+}
We want the semantic information in textual rationales to be task-specific, rather than general purpose representations from pretrained language models, such as BERT \cite{devlin2018bert}. 
Inspired by the pretraining/finetuning paradigm in NLP, 
we propose to treat the rationales ranking task as a downstream task, and finetune\footnote{the finetuning of transformer model serves as a pretrain task in our SE-BPER+} the transformer model on it to get task-specific representations to improve the rationale ranking performance.
We follow the common finetuning setup introduced in the original BERT paper;
the objective function is to maximize
\begin{equation}
    \sum_{u \in \mathcal{U}} \sum_{i \in \mathcal{I}} \sum_{e \in \mathcal{E}} \sum_{e'} f(\textbf{p}_u, \textbf{q}_i, \textbf{e}_{\text{[CLS]}}) - f(\textbf{p}_u, \textbf{q}_i, \textbf{e}'_{\text{[CLS]}})
\end{equation}
where $e'$ are random negative rationales sampled from $\mathcal{E}\backslash(\mathcal{E}_u \bigcup \mathcal{E}_i)$.
\begin{equation}
    f(\textbf{p}_u, \textbf{q}_i, \textbf{e}_{\text{[CLS]}}) = \textit{FFN}([\textbf{p}_u; \textbf{q}_i; \textbf{e}_{\text{[CLS]}}])
\end{equation}
where $[\cdot; \cdot; \cdot]$ denotes the concatenation of three vectors, $\textit{FFN}(\cdot)$ is a single-layer fully connected network with input dimension of $768 \times 3$ and will output a scalar as the preference of $(u,i)$ over rationale $e$. 
The objective function is based on the same intuition of BPR \cite{rendle2012bpr}: to maximize the distance between the ground-truth rationales and the random negative rationales.
By introducing this additional finetuning phase, we adapt the general purpose language model to the task of rationales ranking, which in theory overcomes the problem of static representations in BPER+.

In our experiment, we adopt ALBERT (A Lite BERT) \cite{lan2019albert}, a light weight transformer-based model to reduce the computational overhead. 
The rationale embeddings are updated in training stage, and after each epoch, we update the user embedding $\textbf{p}_u$ and item embedding $\textbf{q}_i$ with Eq. \ref{eq:p_u} and Eq. \ref{eq:q_i}, respectively. 
After the pretraining phase, we encode the rationales with the learned ALBERT model to get the [CLS] token embeddings and use them to initialize the user, item and rationale embeddings.
And we perform the same training procedure as SE-BPER.


\section{Experimental Settings}
\label{sec:experiment}

\subsection{Datasets and Partitions}
We use the same EXTRA datasets as previous works \cite{li2021extra,li2021learning}. EXTRA consists of three datasets collected from different domains: Amazon Movies \& TV\footnote{http://jmcauley.ucsd.edu/data/amazon}, TripAdvisor\footnote{https://www.tripadvisor.com} and Yelp\footnote{https://www.yelp.com/dataset/challenge}. 
Each record in the three datasets consists of user ID, item ID and one or multiple rationale IDs, thus results in one or multiple user-item-rationale triplets. 
The rationales are in the shape of text snippets. 
We show a detailed statistics in Table \ref{tab:dataset} and refer the readers to the original dataset paper \cite{li2021extra} for more details.
To keep the results reproducible and directly comparable, we use the original train-test partitions provided by the EXTRA dataset.
The reported results are mean values from five different runs as the EXTRA dataset provides five different train-test partitions for cross validation.

\begin{table}[t]
\caption{The basic statistics of the datasets}

\resizebox{0.95\columnwidth}{!}{
\begin{tabular}{l|rrr}
\toprule
Dataset                         & Amazon & TripAdvisor & Yelp \\ 
\hline
\# of users                     & 109,121 & 123,374 & 895,729 \\
\# of items                     & 47,113 & 200,475 & 164,779 \\
\# of explanations              & 33,767 & 76,293 & 126,696 \\
\# of $(u,i)$ pairs             & 568,838 & 1,377,605 & 2,608,860 \\
\# of $(u,i,e)$ triplets        & 793,481 & 2,618,340 & 3,875,118 \\
\# of rationales/$(u,i)$ pair & 1.39 & 1.90 & 1.49 \\
Density $(\times 10^{-10})$     & 45.71 & 13.88 & 2.07 \\
\bottomrule
\end{tabular}
}
\label{tab:dataset}
\end{table}

\subsection{Compared Methods}
\begin{table*}[t]
\caption{The personlized rationales ranking performance, we mark the best baseline with $*$, the best performing model with $\dag$, and \textbf{Bold} indicates statistically significant improvement over other models with one-sided t-test at 0.05 level.}
\resizebox{\textwidth}{!}
{
\begin{tabular}{l|cccc|cccc|cccc}
\toprule
Dataset & \multicolumn{4}{c|}{Amazon} & \multicolumn{4}{c|}{TripAdvisor} & \multicolumn{4}{c}{Yelp} \\ 
\hline
Models/Metrics & nDCG@10 & Pre@10 & Rec@10 & F1@10 & nDCG@10 & Pre@10 & Rec@10 & F1@10 & nDCG@10 & Pre@10 & Rec@10 & F1@10 \\
\hline
RAND & 0.014 & 0.004 & 0.027 & 0.006 & 0.005 & 0.002 & 0.011 & 0.004 & 0.003 & 0.001 & 0.007 & 0.002 \\
PITF & 9.366 & 1.824 & 14.125 & 3.149 & 3.725 & 1.111 & 5.851 & 1.788 & 1.844 & 0.635 & 4.172 & 1.068 \\
BPER & 10.285 & 1.942 & $15.147^*$ & $3.360^*$ & 4.091 & 1.236 & 6.549 & 1.992 & 1.912 & 0.723 & 4.768 & 1.218 \\
BPER+ & $10.623^*$ & 1.919 & 14.936 & 3.317 & $4.755^*$ & $1.565^*$ & $8.151^*$ & $2.515^*$ & $1.957^*$ & $0.742^*$ & 4.544 & 1.225 \\ BPER-large & 10.422 & $1.956^*$ & 15.044 & 3.462 & 4.108 & 1.241 & 6.556 & 2.087 & 1.928 & 0.732 & $4.772^*$ & $1.270^*$ \\
\hline
SE-BPER & 11.833 & 2.175 & 16.541 & 3.844 & 5.042 & 1.574 & 8.332 & 2.647 & 2.011 & 0.767 & 4.925 & 1.327 \\
SE-BPER+ & $\textbf{12.415}^{\dag}$ & $\textbf{2.306}^{\dag}$ & $\textbf{17.452}^{\dag}$ & $\textbf{4.074}^{\dag}$ & $5.117^{\dag}$ & $\textbf{1.637}^{\dag}$ & $\textbf{8.560}^{\dag}$ & $\textbf{2.748}^{\dag}$ & $2.024^{\dag}$ & $0.782^{\dag}$ & $4.981^{\dag}$ & $1.352^{\dag}$ \\  
\bottomrule
\end{tabular}

}
\label{tab:results}
\end{table*}

We compare the proposed method with the following algorithms:

\textbf{RAND}: this is a weak baseline that randomly picks up rationales from the rationale collection $\mathcal{E}$. 

\textbf{PITF}: Pairwise Interaction Tensor Factorization \cite{rendle2010pairwise}. It makes prediction for a triplet based on 
\begin{equation}
    \hat{r}_{uie} = \textbf{p}_u^{\text{T}} \textbf{o}_e^U + \textbf{q}_i^{\text{T}} \textbf{o}_e^{I}
\end{equation}
and the embedding vectors are optimized by minimizing
\begin{equation}
    \sum_{u \in \mathcal{U}} \sum_{i \in \mathcal{I}} ( \sum_{i' \in \mathcal{I}\backslash \mathcal{I}_u} - \ln \sigma(\hat{r}_{u,ii'}) + \alpha \sum_{e \in \mathcal{E}_{ui}} \sum_{e' \in \mathcal{E}\backslash \mathcal{E}_{ui}} -\ln \sigma(\hat{r}_{ui,ee'}) )
\end{equation}
where $i'$ and $e'$ are negative samples, $\mathcal{I}_u$ denotes the item set user $u$ has interacted with, $\mathcal{E}_{ui}$ denotes the rationales mentioned by user $u$ on item $i$, and $\alpha$ is a hyperparameter to balance two training objectives.

\textbf{BPER}: Bayesian Personalized Explanation Ranking \cite{li2021learning}. It is a strong baseline to train user, item and rationale embeddings by pairwise loss; however, it fails to learn the semantic information from the textual rationales.

\textbf{BPER+}: BERT-enhanced BPER \cite{li2021learning}. This is an improved version of BPER by incorporating the semantic information encoded by a static transformer-based pretrained language model (BERT in the implementation).

\textbf{BPER-large}: We also study how BPER will perform when we use a latent dimension size of 768, which is the same as SE-BPER and SE-BPER+. We use random initialization and term this ablation as BPER-large.

\subsection{Evaluation and Implementation Details}
\subsubsection{Evaluation Metrics.} We use the same evaluation metrics as in the EXTRA paper \cite{li2021extra} and the BPER paper \cite{li2021learning}. 
More specifically, we use Precision (\textbf{Pre}), Recall (\textbf{Rec}), \textbf{F1} and normalized Discounted Cumulative Gain (\textbf{nDCG}).
nDCG is a measure to evaluate the ranking quality, where $\textit{nDCG}@p = \textit{DCG}@p / \textit{IDCG}@p$, where $p$ denotes cutoff position, 
$$
\textit{DCG}@p = \sum_{i=1}^p \frac{\textit{rel}_i}{\log_2(i+1)}
$$
$$
\textit{IDCG}@p = \sum_{i=1}^{\mathcal{E}_{ui}} \frac{\textit{rel}_i}{\log_2(i+1)}
$$
here $\mathcal{E}_{ui}$ denotes the rationales for the $(u,i)$ pair in the testset and
$\textit{rel}_i \in \lbrace 0,1 \rbrace$ denotes whether the rationale is in $\mathcal{E}_{ui}$ or not\footnote{the nDCG results we report are different from Li et al. \cite{li2021learning} because they use a different nDCG function}.
To keep the results directly comparable, we evaluate the same top-10 ranking as \cite{li2021learning}. 

\subsubsection{Parameters and Hyperparameters.}
For SE-BPER and SE-BPER+, we optimize the parameters with Adam \cite{kingma2014adam}; we search the learning rate from 1e-1 to 1e-5, L2 normalization from 1e-1 to 1e-5 and the number of negatives in Eq. \ref{eq:loss} from 2 to 5. We set the $\mu$ hyperparameter in Eq. \ref{eq:score} to 0.7, which is the same as the original BPER implementation. In SE-BPER+'s pretrain, we set the learning rate to 5e-6 for ALBERT model and 1e-4 for the single-layer FFN.

\subsubsection{Implementation.} 
We implement our SE-BPER and SE-BPER+ with PyTorch\footnote{https://pytorch.org/} and albert-base-v2 model\footnote{albert-base has 11M parameters compared to bert-base's 110M parameters} from the HuggingFace transformer library\footnote{https://huggingface.co/}. 
For baselines, we implement RAND and PITF by PyTorch, and we use the official implementation\footnote{https://github.com/lileipisces/BPER} of BPER and BPER+.

\section{Results and Analysis}
\label{sec:result}
In this section, we study the effectiveness of the proposed algorithms. Specifically, we are interested in two research questions: \\
\textbf{RQ1}: How do SE-BPER and SE-BPER+ perform in the rationales ranking task compared to BPER? \\
\textbf{RQ2}: Can the good initialization strategy, i.e. initialization with semantic embeddings speed up the model convergence?

\subsection{Performance Comparison (RQ1)}
We report the rationales ranking performance in Table \ref{tab:results}. 

\subsubsection{Comparison between baselines:}
Among the baselines, BPER+ generally achieves better performance. For example, BPER+ has better nDCG in Amazon dataset, better nDCG, Precision, Recall and F1 in
TripAdvisor dataset, and better nDCG, Precision in Yelp dataset. However, it is also outperformed by BPER and BPER-large in some cases. For example, BPER-large has better Recall and F1 in Yelp dataset.
There might be two reasons behind BPER+'s inconsistent performance:
(1) the meaningful semantic information is lost in the projection phase as in Eq. \ref{eq:projection}. (2) in Eq. \ref{eq: u_mul} and Eq. \ref{eq: i_mul}, the design choice of multiplying two vectors is not the optimal way to combine two sources of information.

\subsubsection{Study of SE-BPER's performance.}
From Table \ref{tab:results} we can see SE-BPER consistently outperforms the baselines\footnote{all the reported improvement numbers are absolute improvement}. For example, in the Amazon dataset, SE-BPER outperforms BPER+ by 1.210\% in nDCG, 0.256\% in Precision, and 1.394\% in Recall; in the TripAdvisor dataset, the improvements are 0.275\%, 0.009\%, 0.171\%, respectively. 
This supports our argument that compared to BPER+, the proposed SE-BPER can better combine the information from interaction matrix and the semantic information within the textual rationales.
We also notice that SE-BPER is consistently better than BPER, this echoes the argument from \cite{li2021extra} that adding textual information from rationales can help with the rationales ranking performance.

We also add a specific ablation study, i.e. BPER-large to see whether the improvement of performance is brought by the increase of embedding dimensions. 
We notice that BPER-large outperforms the vanilla BPER by a small margin, but the improvement is not significant. 
SE-BPER still achieves significant better performance compared to BPER-large, which suggests the effectiveness of the proposed algorithm.

\subsubsection{Comparison between SE-BPER and SE-BPER+.}
We are interested to know whether adding a pretraining stage can help our BERT-style transformer model generate task-specific semantic representations to improve the downstream task's performance (rationale ranking in our case). From Table \ref{tab:results} we can see that SE-BPER consistently outperforms SE-BPER (and all the baselines). For example, in the Amazon dataset, the improvments are 0.582\%, 0.131\%, 0.911\% and 0.230\% in nDCG, Precision, Recall and F1, respectively, and all improvements passed the paired t-test with confidence level at 0.01. This suggests that adding a task-specific pretraining phase could indeed help with the model performance. 

To summarize from Table \ref{tab:results}, the proposed SE-BPER outperforms the baselines, and adding a pretraining stage, i.e. SE-BPER+ can further boost the performance.

\begin{table}[t]
\caption{Training time required for model convergence; h denotes hours; all three models share the same embedding dimension of 768. We stop the model training after the performance hasn't been improved for 3 consecutive epochs and record the best performing epoch. We highlight the minimum amount of training time. All numbers are reported with batch size of 32, learning rate of 1e-3 and L2 regulargization coefficient of 1e-4}
\resizebox{0.8\columnwidth}{!}{
\begin{tabular}{l|ccc} \toprule
Model/Dataset & Amazon & TripAdvisor & Yelp \\ \hline 
BPER-large & 4.0 h & 18.9 h & 31.8 h \\
SE-BPER & 2.0 h & \textbf{10.8 h} & \textbf{21.2 h} \\
SE-BPER+ & \textbf{1.6 h} & \textbf{10.8 h} & \textbf{21.2 h} \\

\bottomrule
\end{tabular} }
\label{tab:converge}
\end{table}

\subsection{Convergence Speed (RQ2)}

Previous works \cite{boutsidis2008svd,hidasi2012enhancing,hidasi2013initializing} also suggest that better initialization strategy for latent factor models can speed up the model convergence. 
We report the model training time\footnote{the reported running time are all mean values from five different runs} required for BPER-large, SE-BPER and SE-BPER+'s convergence in Table \ref{tab:converge}; 
All experiments are conducted on a Linux server with AMD Ryzen Threadripper PRO 3955WX 16-Cores CPU (3.9G Hz), a single NVIDIA RTX A500 GPU (24 GB) and 128 GB memory size.
We don't include the baselines as their embedding dimensions are much smaller thus not suitable for comparison;
and we don't report the inference time as the compared models share the same inference time because the only differences are the initialization strategies.

From Table \ref{tab:converge}
we can see that compared to random initialization, i.e. BPER-large, the proposed SE-BPER achieves faster convergence in all three datasets with the same learning rate and L2 regularization coefficient. 
For example, SE-BPER requires approximate 1.6 hours to converge in Amazon dataset compared to random initialization, i.e. BPER-large's 4.0 hours.
SE-BPER+ further has faster convergence in Amazon dataset, but the improvement is not significant. 
This suggests that utilizing semantic representation can speed up the model convergence.


\section{Conclusions}
\label{sec:conclusion}
Within the scope of this work, we study the task of rationales ranking in model-agnostic explainable recommendation. 
We argue that the SOTA algorithm BPER+ fails to effectively utilize the semantic information in the candidate textual rationales. 
We propose Semantic-Enhanced Bayesian Personalized Explanation Ranking (SE-BPER) to combine the semantic information from textual rationales and the information from interaction matrices. 
Extensive experiments with ablation studies prove the superiority of SE-BPER in ranking performance and model convergence. 
We also propose a pretrain phase to build task-specific representations for SE-BPER, which we term as SE-BPER+. SE-BPER+ can further boost the performance of SE-BPER. 
We note that rationale ranking falls into the scope of personalized ranking, where the optimal way to combine textual and interaction information remains to be discovered.

\section{Acknowledgement}
This work was supported in part by the School of Computing, University of Utah and in part by NSF IIS-2007398. Any opinions, findings and conclusions or recommendations expressed in this material are those of the authors and do not necessarily reflect those of the sponsor.

\bibliographystyle{ACM-Reference-Format}
\balance
\bibliography{reference}

\end{document}